\documentclass[aip,apl,twocolumn,groupedaddress,showpacs,showkeys]{revtex4}
\usepackage{amsmath}
\usepackage{amssymb}
\usepackage[load-configurations=abbreviations,separate-uncertainty=true]{siunitx}
\usepackage{graphicx}

\usepackage{xcolor}

\begin{document}

% *************************
% ***   head of paper   ***
% *************************
\title{Radiation induced electronic trap states and local structural disorder in van~der~Waals bonded semiconductor crystals}

\author{Tobias~Morf}
\email[]{tmorf@phys.ethz.ch}

\author{Tino~Zimmerling}
\email[]{tino.zimmerling@phys.ethz.ch}

\author{Simon~Haas}
\email[]{s.haas@alumni.ethz.ch}

\author{Bertram~Batlogg}
\email[]{batlogg@phys.ethz.ch}
\affiliation{Laboratory for Solid State Physics, ETH Zurich, 8093 Zurich, Switzerland}

\date{\today}

\begin{abstract}
	In controlled X-ray irradiation experiments, the formation of trap 
	states in the prototypical van~der~Waals bonded semiconductor 
	Rubrene is studied quantitatively for doses up to \SI{82}{Gray}~%
	($\si{\gray}=\si{\joule\per\kg}$). About \num{100} electronic 
	trap states, located around \SI{0.3}{\eV} above the valence 
	band, are created by each absorbed \SI{8}{\keV} photon 
	which is 2--3 orders of magnitude more than \SI{1}{\MeV} 
	protons produce. Thermal annealing is shown to reduce these traps.
	Local structural disorder, which has also been induced by other means 
	in different studies, 
	is thus identified as a common origin 
	of trap states in van~der~Waals bonded molecular organic semiconductors.
\end{abstract}

	% Conductivity of specific materials: Polymers; organic compounds (including organic semiconductors) 


\maketitle
% *************************
% ***   body of paper   ***
% *************************

The understanding of charge traps in van~der~Waals bonded semiconductors has 
reached new levels in recent years and organic electronic 
devices' performance has much improved with promising perspectives 
for applications.
With the emerging understanding and spectral analysis of the trap 
density of states (DOS) \cite{Salleo2013,Kalb2010b} 
it is highly desirable to quantify the interaction of environmental 
influences with van~der~Waals bonded semiconductors in terms of density and spectral 
distribution of the induced trap states. Such environmental influences 
may occur during fabrication, storage or operation of an organic 
electronic device. They cover a broad range of chemical \cite{Krellner2007}, 
mechanical \cite{Sekitani2005} and also radiative \cite{Quaranta2003,Newman2007,Zimmerling2012} 
phenomena.

The commonly known adverse influence of traps on charge transport 
\cite{Kalb2010b} as well as spin diffusion length \cite{Rybicki2012} 
is contrasted by an unexpected positive influence of X-ray radiation 
in electron-beam evaporation processes on the magnetotransport in 
organic materials \cite{Rybicki2012}. These observations 
are a strong motivation to further investigate the defects arising 
from X-ray radiation.

Furthermore, after intense scintillation studies in the 1970s 
\cite{Weisz1966,Birks1951b,Yokoi1978,Shiomi1967}, 
organic materials are considered anew for direct X-ray detection 
through the photoconductivity effect \cite{Newman2007, Blakesley2007}. 
For applications in low-cost, large-area integrated X-ray imaging 
panels it will be of central importance to assess the radiation 
damage and whether those defects can be healed by thermal 
annealing.

In this study we address the formation of electronic trap 
states upon X-ray irradiation in Rubrene single crystals, a 
grain boundary free model material for van~der~Waals bonded semiconductors. 
The spectral density of trap states (DOS) is determined 
by measuring current voltage 
characteristics at different temperatures and applying temperature-dependent 
space-charge limited current spectroscopy (TD-SCLC). The basic 
concept of SCLC is electrical transport by charge carriers thermally 
excited above a certain energy separating extended from localised states. 
No further \emph{a priori} assumptions --- in particular no 
specific transport model --- are required. Due to the Fermi-Dirac statistics this excitation from 
localised traps to delocalised conducting states takes place in a 
small energy window. With increasing voltage, more space charge is 
injected, hence the Fermi energy $E_F$ is shifted towards the delocalised states. The trap 
DOS is calculated from this differential increment. The energy 
scale is given by the thermal activation energy at a given voltage 
corrected by the statistical shift, which accounts for the asymmetry 
of the DOS around $E_F$. The full procedure is formally discussed 
in references \onlinecite{Zmeskal1985,Schauer1986,Schauer1996,Schauer1997,Nespurek1984,Braga2008} 
and numerically implemented using cubic smoothing splines \cite{deBoor1987}.

\begin{figure}[b]
	\includegraphics{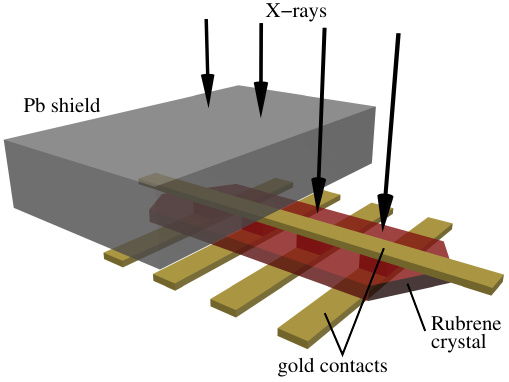}%
	\caption{(Colour online) Schematic of the experiment. The 
		Rubrene crystal is laminated onto prefabricated 
		bottom electrodes, then the top electrode is evaporated. 
		The lead shield screens part of the crystal from X-rays, 
		thus allowing direct comparison of irradiated to unirradiated 
		crystal sites.\label{fig:SampleLayout}}
\end{figure}

Rubrene crystals were grown by physical vapour transport 
in high purity argon flow. The platelet-like crystals were then 
laminated onto prefabricated gold electrodes, similar to the 
`flip-crystal' technique \cite{Takeya2003}. For the vacuum deposition 
of the top electrode, the samples were cooled in order to minimise 
the thermal load on the crystals. The sample layout is schematically 
shown in Fig.~\ref{fig:SampleLayout}. Current flows along the 
\SI{26.86}{\angstrom} long crystallographic $a$ axis \cite{Jurchescu2006,Menard2006,Minato2009}.
The simultaneous measurement of up to four sites on the same 
crystal --- called channels --- provides verifiable results and 
the shielding of some of these channels during irradiation provides 
the necessary reference.
Furthermore, this sample 
arrangement (c.f.\ Fig.~\ref{fig:SampleLayout}) enables checking 
of reproducibility and device stability.

The SCLC measurements were performed in 
darkness in a cryostat's helium atmosphere. Charge 
carrier injection from the laminated bottom electrodes turned out 
to be more efficient than from the deposited top electrode and 
thus the polarity for all measurements was chosen accordingly. 
Current and power limits prevent crystal damage \cite{Srour1998,Srour2009} 
or local heating.

For the quantitative study of the radiation damage, a crystal 
diffractometer served as a well-defined monochromatic CuK$\alpha$ 
(\SI{8}{\keV}) radiation source.
The lateral intensity distribution (beam profile) was 
measured with the diffractometer's image-plate detector and 
suitable Zirconium attenuators.
For the measurement of the intensity in absolute units, a suitably 
calibrated instrument was 
kindly provided by the University Hospital Zurich. 
The beam's peak intensity was \SI{3}{\micro\W\per\cm\squared} 
corresponding to a photon flux of \SI{2e9}{\per\s\per\cm\squared}. 
Approximating Rubrene as a 42:28 mixture of carbon and hydrogen 
the crystal at the center of the beam absorbed a dose of 
\SI{43(9)}{\gray} ($=\si{\joule\per\kg}$)~\cite{XrayCoef} 
during one hour of exposure. The \SI{30}{\nm} thick gold top 
contact absorbs only about \SI{1}{\percent} of the incident 
intensity. For comparison, a single computed tomography (CT) scan 
accounts for up to \SI{10}{\milli\gray}~\cite{Brenner2007}, 
typical radiotherapy doses are some \SI{10}{\gray}~\cite{KrayenbuhlPrivCom} 
and the accumulated lifetime dose of X-ray imaging sensors is a 
few \SI{100}{\gray} \cite{Blakesley2007}.

After first measuring the trap DOS of the pristine crystals, the 
samples were transfered to an Argon filled glass tube with a Kapton 
window and aligned in the X-ray beam with fluorescent marks. During 
irradiation, two out of four channels were 
shielded by a \SI{0.1}{\mm} lead foil thus providing reference 
data on the same crystal. After each hour of irradiation, the 
samples were measured and the trap DOS calculated. These repeated 
measurements required the crystals to be stable over multiple 
thermal cycles between \SI{300}{\K} and \SI{100}{\K}.

\begin{figure}
	\includegraphics{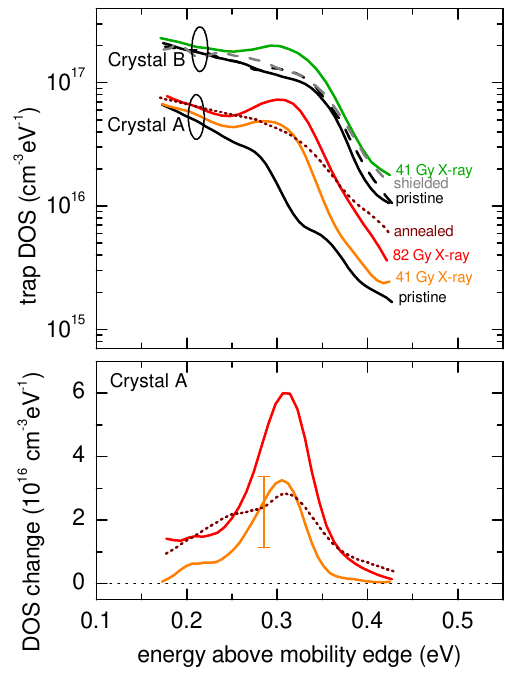}%
	\caption{(Colour online) Trap density of states and its change upon 
		X-ray irradiation in two different Rubrene single crystals. The mobility 
		edge is chosen as the energy reference point. The distinct 
		increase peaked at \SI{0.3}{\eV} is attributed to X-ray induced 
		defects since the shielded part (example shown in dashed grey in upper panel) 
		does not change significantly. Annealing of the sample (dotted line) 
		suggests that structural defects contribute to the total trap state density.\label{fig:DOS}}
\end{figure}

The densities of states measured on two crystals before and after irradiation 
are shown in Fig.~\ref{fig:DOS}. 
The unirradiated channels do not show any significant change in the 
trap DOS compared to the pristine state --- an example is shown for 
crystal B. Thus, X-ray induced defects are clearly identified. 
Furthermore, the unchanged DOS in unirradiated channels reflects the 
stability and reproducibility of sample handling and measurement.

In the exposed channels, the trap DOS increases by up to 
\SI{8e16}{\per\cubic\cm\per\eV} (Fig.~\ref{fig:DOS}, crystal B) 
in a narrow energy range peaked around \SI{0.3}{\eV}. The area 
under this peak yields the total trap density of 
\SI{\sim 3e15}{\per\cubic\cm} traps generated during one hour of 
X-ray irradiation. After the second hour of irradiation, the induced 
trap density has doubled within experimental uncertainty. The X-ray 
absorption length $\lambda \approx \SI{2}{\mm}$ is much 
longer than the typical crystal thickness of \SI{\sim 1}{\um} (only 
\SI{0.05}{\percent} of the photons are absorbed), and with an 
hourly dose of \SI{43}{\gray}, approximately \num{4e13} photons 
are abosrbed per \si{\cubic\cm} creating a uniform defect density.

\begin{figure}
	\setlength{\unitlength}{1mm}
	\begin{picture}(86,74)(0,0)
		\put(0,0){\includegraphics{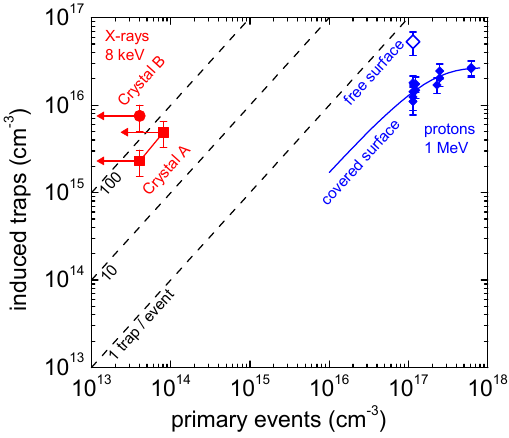}}
		\put(33,15){\resizebox{20\unitlength}{!}{\includegraphics{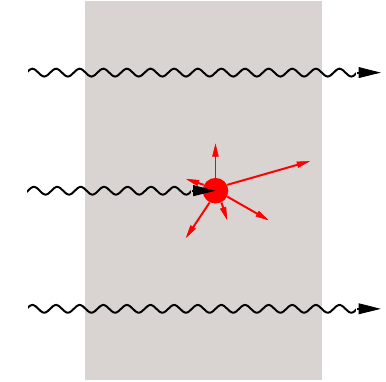}}}
		\put(60,15){\resizebox{20\unitlength}{!}{\includegraphics{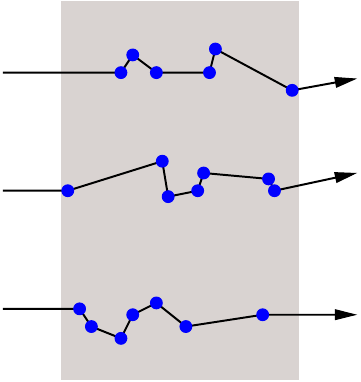}}}
		\put(61.5,37){\makebox[\width][l]{\scriptsize\sffamily Rubrene crystal}}
	\end{picture}
	\caption{(Colour online) Trap density in Rubrene crystals irradiated 
		by X-rays (red) or protons (blue, \cite{Zimmerling2012}). For each 
		primary interaction event, protons create \SIrange{\sim0.2}{0.5}{traps}, 
		while X-rays produce \SI{\sim e2}{traps}. The \numrange{100}{1000} times 
		higher trap creation efficiency of X-rays is attributed to secondary 
		events.	Trap creation by ions in covered crystals saturates 
		due to re-attachement of hydrogen. The insets schematically show 
		the energy deposition processes. An X-ray photon will either pass 
		the crystal undisturbed or deposit its full energy in a single 
		event creating numerous secondary events. On the other hand, 
		every ion will experience several interactions with (mainly) 
		target electrons every time depositing a fraction of its initial 
		energy.
		\label{fig:DoseResponse}}
\end{figure}

To quantitatively compare X-ray and ion irradiation in Fig.~\ref{fig:DoseResponse} 
it is appropriate to consider the microscopic interaction mechanism 
as sketched in the insets. An absorbed X-ray photon will deposit its 
full energy of \SI{8}{\keV} in a single \emph{primary event} causing 
a cascade of secondary events which in turn create numerous microscopic 
defects. In contrast, every proton of \SI{1}{\MeV} experiences 
approximately \num{14} \emph{primary interactions} on its way 
through a \SI{1}{\um} thick crystal, each time transfering 
\SI{\approx2.3}{\keV} to the crystal's electronic system \cite{Zimmerling2012,Pope1999}. 
Simulations using the same SRIM \cite{Ziegler1985} parameters as in ref.~\onlinecite{Zimmerling2012} 
show that atom displacement is negligible.

The electronic excitations allow for hydrogen atoms to be detached 
from Rubrene molecules \cite{Barillon2003}. They will then diffuse through the crystal 
and cause structural disorder as interstitials in the Rubrene lattice. 
Since the crystal surfaces in this study were not covered during irradiation, 
it was possible for detached hydrogen to escape from the crystal. 
It is thus appropriate to consider the open surface data from reference 
\onlinecite{Zimmerling2012} (open symbols in Fig.~\ref{fig:DoseResponse}) 
for a comparison. 

The density of radiation-induced traps is plotted as a function of 
primary events in Fig.~\ref{fig:DoseResponse}. Note that the 
primary event counts of the X-ray datapoints are an upper limit 
based on the assumption that the samples were centred at peak 
intensity. Solid blue (proton radiation) symbols 
are data from covered surfaces saturating at high dose due to 
re-attachement of hydrogen knocked off in a previous event. For 
each primary interaction event, protons create \SI{\sim 0.5}{traps}, 
while X-rays produce \SI{\sim 100}{traps}. A central result of 
this study is: per primary interaction event, X-rays are found to 
be \numrange{100}{1000} times more effective than ions in trap 
generation. This is attributed to the shower of secondary events 
following every photon absorption. These showers spatially distribute 
the energy in the crystal as opposed to the point-like event of an 
ion interaction. These secondary events apparently carry enough 
energy to create defects and their number would account for the 
100- to 1000-fold defect creation rate.

\begin{figure}
	\setlength{\unitlength}{1mm}
	\begin{picture}(86,59)(0,0)
		\put(0,0){\includegraphics[width=\columnwidth]{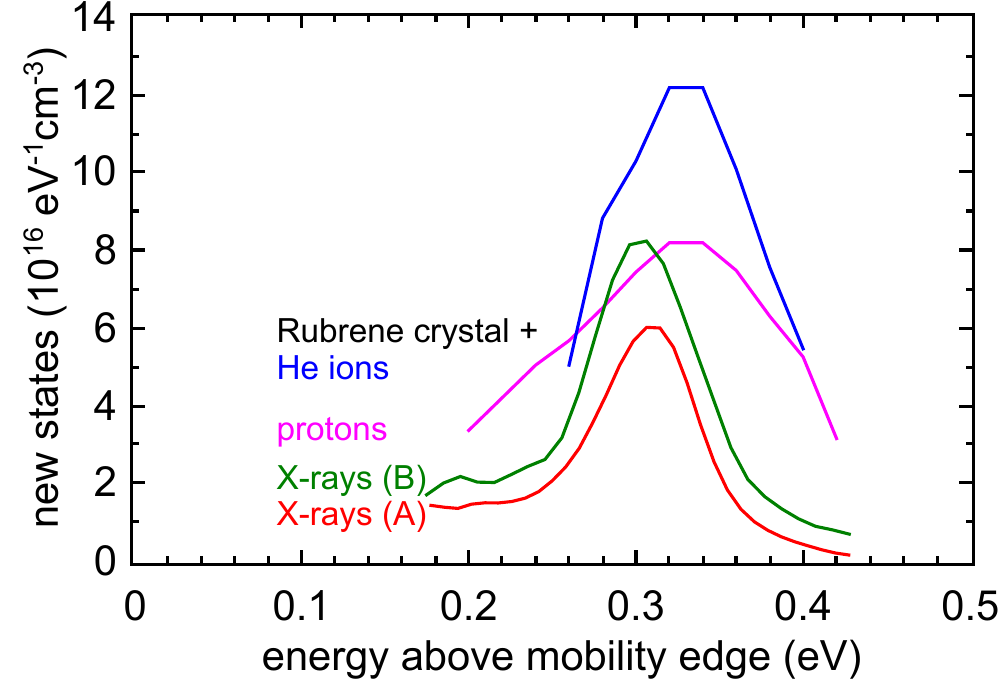}}
		\put(15,38){\includegraphics[width=0.35\columnwidth]{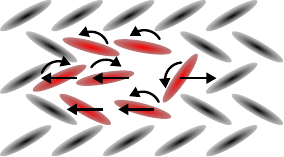}}
	\end{picture}
	\caption{(Colour online) Spectral distribution of disorder induced states in 
		Rubrene single crystals after Helium ion irradiation (blue, \cite{Zimmerling2012}), 
		proton irradiation (magenta, \cite{Zimmerling2012}) and X-ray exposure 
		(green and red, this study). The inset shows an exagerated sketch of local 
		disorder as the possible origin of these traps. \label{fig:peaks}}
\end{figure}

The spectral distribution of the additional traps suggests their common 
microscopic origin: they are peaked \SI{\approx 0.3}{\eV} above 
the (hole) mobility edge and the peak is \SI{\approx 0.1}{\eV} 
wide. The summary in Fig.~\ref{fig:peaks} shows data for Rubrene 
crystals irradiated with protons or Helium ions \cite{Zimmerling2012}, 
together with X-ray data from this study. Most remarkably, a very similar 
position and distribution width has been found in Pentacene thin films 
exposed to oxygen \cite{Kalb2008} and in UV/ozone exposed Rubrene 
single crystals \cite{Krellner2007}. 
Furthermore, recent low background UPS studies \cite{Bussolotti2013} 
also report the formation of energetically very similar traps 
when oxygen but also chemically neutral nitrogen or argon penetrate Pentacene films and the 
same defects have been found in both photocurrent measurements in 
organic solar cells\cite{Street2012} and density functional calculations of specific 
hydrogen- and oxygen-related defects\cite{Northrup2003,Northrup2013}.

To further elucidate the origin of radiation induced trap 
states, Crystal A in Fig.~\ref{fig:DOS} has been annealed for 
\SI{12}{\hour} at \SI{350}{\K} in helium atmosphere. Most significantly, a reduction 
of radiation defects by approximately \SI{40}{\percent} is observed.
This is a central result and is in line with previous observations 
in anthracene crystals \cite{Heppell1967,Yokoi1978,Zorn1993}. 

A motivation to consider structural defects as a possible cause is 
the partial recovery after thermal annealing at moderate temperatures. 
Annealing of structural defects even at room temperature has been 
shown to take place in Pentacene thin films always kept in high 
vacuum \cite{Kalb2007}. Similarly, intercalation with inert gases 
(N$_2$ and Ar) induces trap states at a similar energy \cite{Bussolotti2013}. 
Also in organic polymer solar cells, radiation induced damage 
recovering by annealing has been observed and interpreted in terms 
of hydrogen detachment and rearrangement \cite{Street2012,Northrup2013,Nakagawa1976}. 
Particularly interesting is the bending of a pentacene molecule when 
two hydrogen atoms are attached to the initially flat entity, creating 
localised electronic states \cite{Northrup2003}. Similar detailed 
calculations for Rubrene \cite{Tsetseris2008} with attached oxygen 
of hydrogen in various configurations give no evidence for new 
electronic states within the few tenths of \si{\eV} above the HOMO 
band accessible in the experiment. However, they reveal a slight 
rotation of a phenyl side-group. While the authors are not aware 
of a corresponding calculation involving a missing hydrogen in Rubrene, 
the previous experimental observations \cite{Zimmerling2012} clearly 
suggest hydrogen detachment to be a key step in trap state formation.

The discussion about the microscopic nature of intentionally induced 
defect states in organic semiconductor crystals is now stimulated by 
(a) the formation of electronically active states \SI{\approx 0.3}{\eV} 
above the HOMO band and (b) the ability to partially anneal them at 
very moderate temperatures. Future studies therefore might address 
the relative impact of new chemical species  perturbing the $\pi$ electron 
system (e.g.\ H detachment, O or OH attachement), and 
on modifications of the molecule's structure and its environment in 
the crystal. Combining local probes and macroscopic transport 
measurements will produce such new insights.

In conclusion, we have quantitatively studied the formation of bulk 
trap states in van~der~Waals bonded single crystals by X-ray irradiation. 
For each absorbed \SI{8}{\keV} photon, approximately \num{100} trap 
states are created, while proton irradiation \cite{Zimmerling2012} 
generates up to \num{1} trap state per primary interaction.
The spectral trap distribution is peaked near \SI{0.3}{\eV} above 
the HOMO transport level and is \SI{\approx .15}{\eV} wide.
Very similar trap distributions which can be partially annealed are 
produced by hydrogen- and oxygen-related chemical defects but also 
when van~der~Waals bonded 
semiconductors are locally disturbed by proton or Helium ion 
radiation \cite{Zimmerling2012} or by penetration of oxygen or 
chemically neutral nitrogen or argon \cite{Krellner2007,Kalb2008,Kalb2010b,Bussolotti2013}. 
This formation of energetically very similar trap states by a wide 
range of treatments and the observation of partial annealing of these 
states set the framework for future studies focussing in the respective 
contributions of chemical and structural defects.

\begin{acknowledgments}
The authors thank Carlo Bernasconi from the Laboratory of 
Crystallography, ETH Zurich for access to the X-ray diffractometer. 
Stephan Kl{\"o}ck and J{\'e}r{\^o}me Krayenb{\"u}hl from University 
Hospital Zurich are greatfully acknowledged for their help in 
measuring the X-ray intensity and Kurt Mattenberger for support with 
various technical issues.
\end{acknowledgments}

\bibliography{XrayTraps_refs}

\end{document}